\documentclass[aps,prb,showpacs,reprint,amsmath,amssymb,superscriptaddress]{revtex4-1}
\usepackage{graphicx}
\usepackage{dcolumn}
\usepackage{bm}
\usepackage{color}

\begin{document}
\title{Direct band gap carbon superlattices with efficient optical transition}
\author{Young Jun Oh}
\altaffiliation[Present address: ]{Department of Materials Science and Engineering, University of Texas at Dallas, Richardson, TX 75080, USA.}
\affiliation{Department of Physics, Korea  Advanced Institute of Science and Technology, Daejeon 34141, Korea}
\author{Sunghyun Kim}
\affiliation{Department of Physics, Korea  Advanced Institute of Science and Technology, Daejeon 34141, Korea}
\author{In-Ho Lee}
\affiliation{Korea Research Institute of Standards and Science, Daejeon 34113, Korea}
\affiliation{Center for In Silico Protein Science, School of Computational Science, Korea Institute for Advanced Study, Seoul 02455, Korea}
\author{Jooyoung Lee}\email[Corresponding author: ]{jlee@kias.re.kr}
\affiliation{Center for In Silico Protein Science, School of Computational Science, Korea Institute for Advanced Study, Seoul 02455, Korea}
\author{K. J. Chang}\email[Corresponding author: ]{kjchang@kaist.ac.kr}
\affiliation{Department of Physics, Korea  Advanced Institute of Science and Technology, Daejeon 34141, Korea}
\date{\today}

\begin{abstract}
We report pure carbon-based superlattices that exhibit direct band gaps and excellent optical absorption and emission properties at the threshold energy.
The structures are nearly identical to that of cubic diamond except that defective layers characterized by five- and seven-membered rings are intercalated in the diamond lattice.
The direct band gaps lie in the range of 5.6$\sim$5.9 eV, corresponding to wavelengths of 210$\sim$221 nm.
The dipole matrix elements of direct optical transition are comparable to that of GaN, suggesting that the superlattices are promising materials as an efficient deep ultraviolet light emitter.
Molecular dynamics simulations show that the superlattices are thermally stable even at a high temperature of 2000 K.
We provide a possible route to the synthesis of superlattices through wafer bonding of diamond (100) surfaces.
~\\
\end{abstract}
~\\
\maketitle
\section{Introduction} 
Carbon (C) exhibits various allotropes such as diamond, graphite, graphene, fullerene, and nanotubes \cite{C-rev1}.
Among these allotropes, diamond, a metastable crystalline form of carbon, belongs to wide band gap semiconductors such as SiC and GaN.
In addition, diamond exhibits high electric breakdown field, high thermal conductivity, high saturated electron drift velocity, high carrier mobility, and excellent thermal stability \cite{diamond}.
Due to superior electrical and thermal properties, as compared to Si and other wide band gap semiconductors, diamond is considered as a promising material for devices which operate at very high voltages, frequencies, and temperature.
However, diamond is not an efficient light emitter due to its indirect band gap nature.
On the other hand, one-dimensional carbon nanotubes exhibit a variety of electronic properties ranging from metallic to semiconducting \cite{CNT-rev}, while graphene, a two-dimensional allotrope of carbon, has a band gap of zero.
Although light emitting devices were realized based on single-walled carbon nanotubes with direct gaps \cite{CNT-app1,CNT-app2}, their utility as a light source is limited by their nanometre scale size.
 
Recently, a great deal of attention has been paid to the prediction/discovery of new carbon allotropes.
The process of cold compression of graphite \cite{graexp1,graexp2,graexp3,graexp4,graexp5,graexp7} and carbon nanotubes \cite{cntexp1,cntexp2,cntexp3} is known to produce many unknown phases of carbon, and extensive computational studies have been attempted to understand them in terms of their structural clarification.
Numerous carbon allotropes were suggested as candidates, such as M carbon \cite{M2009}, W carbon \cite{W2011}, body-centered tetragonal bct-C$_{4}$ carbon \cite{bct1,bct4}, Z carbon \cite{Z2012} (also named $o$C16-II \cite{oC16II-2011} and Cco-C8 \cite{Cco-C8-2011}),
R carbon \cite{P2012} (also named O carbon \cite{O2012} and H carbon \cite{H2012}), P carbon \cite{P2012} (also named Z4-A$_3$B$_1$ carbon \cite{Z4A3B1-2012}), S carbon \cite{H2012} (also named C carbon \cite{C2012}), F carbon \cite{F2012} (also named J carbon \cite{J2012}), $o$C32 carbon \cite{oC32-2013}, and M585 carbon \cite{M585-2014}.
These allotropes form $sp^{3}$ bonds and their structures are characterized by the topological arrangement of 4-, 5-, 6-, 7-, and 8-membered carbon rings \cite{P2012,Topology-2013}, different from the cubic or hexagonal (lonsdaleite) phase of carbon.

While most carbon allotropes are of indirect band gaps, some allotropes were proposed to be direct band gap semiconductors, for example, S carbon \cite{H2012}, Z-ACA \cite{Z4A3B1-2012}, Z-CACB \cite{Z4A3B1-2012}, C carbon \cite{C2012}, fullerite C$_{24}$ \cite{C24-2006}, T carbon \cite{T2011}, TY carbon \cite{TY2012}, $sp^2$-diamond \cite{sp2-2013}, and $Amm$2 carbon \cite{Amm2-2014}.
Although extensive studies were performed for the structural properties such as energetics, hardness, and phase transition, the electronic properties were not as well studied.
Semiconductors which can emit photons with wavelengths shorter than 350 nm have potential to be used in deep ultraviolet (UV) light emitting diodes (LEDs) and laser diodes (LDs).
Currently, AlGaN alloys have been considered as candidates for deep UV light sources.
Although GaN-based LEDs in the near UV range (at 365 nm) were successfully demonstrated \cite{365nm}, it is still a challenging issue to obtain high external quantum efficiencies in AlGaN-based LEDs with wavelengths below 250 nm \cite{uv3,uv4,led1}.
For applications to LEDs and LDs in a deep UV wavelength range, it is desirable to search for direct band gap carbon allotropes with the gap sizes comparable to that of diamond.

In this work, we find direct band gap carbon allotropes in the form of superlattices through computational searches, based on a combined approach of global optimization and first-principles density functional calculations.
The direct band gaps range from 5.6 to 5.9 eV, suitable for high power applications.
The optical transitions at the threshold energy are comparable to that of GaN, suggesting that the designed allotropes can serve as promising materials for a deep UV light source.

\section{Calculation method}
We explored carbon allotropes with direct band gaps by using a combined computational approach \cite{Lee2014} of conformational space annealing (CSA) \cite{csa} for global optimization and first-principles electronic structure calculations.
We used a variant of CSA where the enthalpy of the system was minimized and the objective function was designed to promote the formation of a direct band gap.
The enthalpy minimization and the analysis of electronic properties were performed within the framework of density functional theory.
We used the functional form of Perdew, Burke, and Ernzerhof (PBE) \cite{pbe} for the exchange-correlation potential and the projector augmented wave potentials \cite{paw}, as implemented in the VASP code \cite{vasp}.
The wave functions were expanded in plane waves with an energy cutoff of 600 eV, and an even higher energy cutoff of 800 eV was used at the final stage of optimization.
Without using any knowledge of known crystal structures of carbon, we optimized the degrees of freedom including atomic positions and lattice parameters for C systems with $N$ atoms per unit cell ($N$ = 8, 12, 16, 20).
The number of configurations was set to 50 in the population size of CSA.

\section{Results and discussion}
\begin{figure*}
\includegraphics[width=\textwidth]{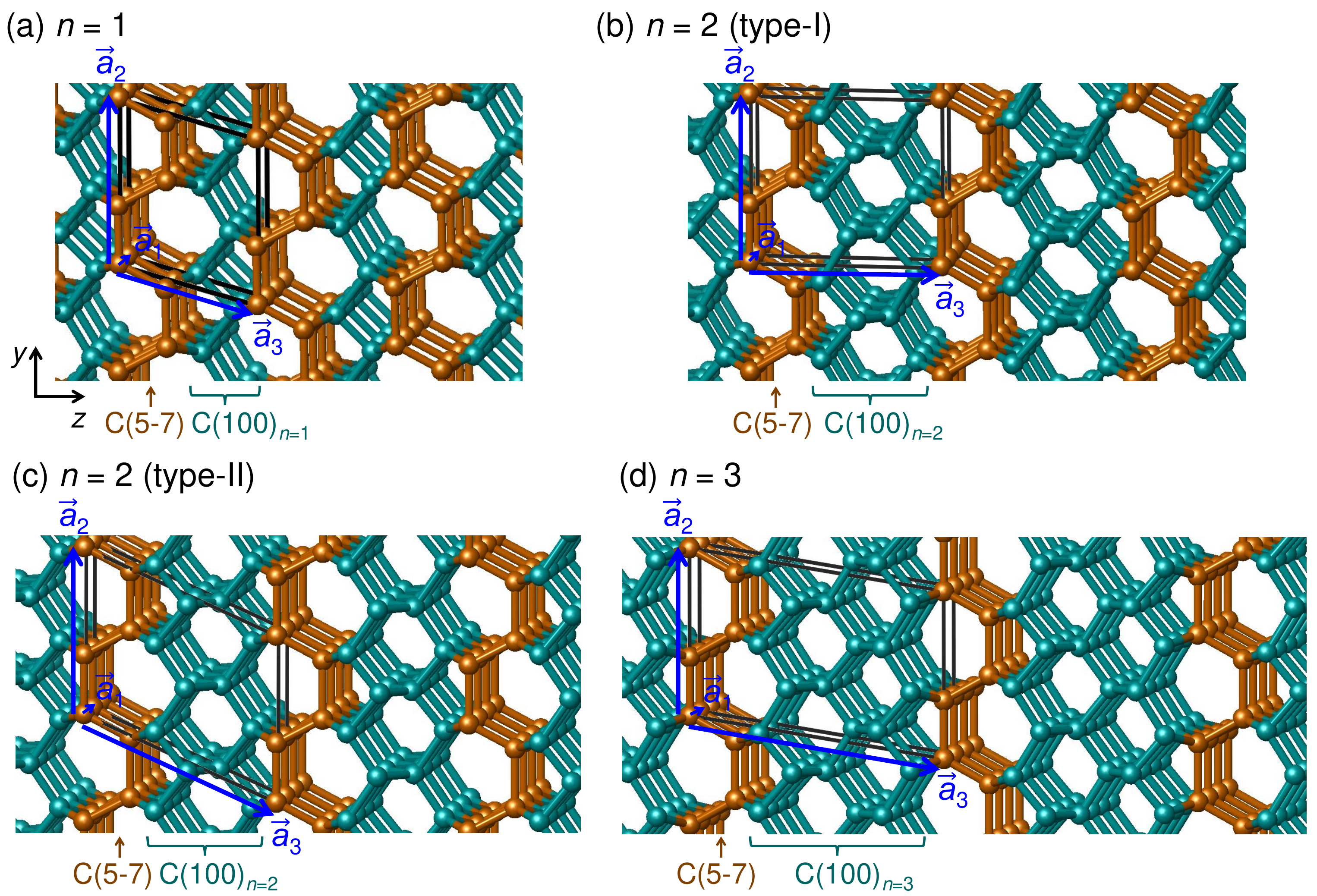}
\caption{The atomic structures of the C(100)$_n$/C(5-7) superlattices for (a) $n$ = 1, (b) $n$ = 2 (type-I), (c) $n$ = 2 (type-II), and (d) $n$ = 3. Colored cyan and brown circles denote the C atoms which belong to the C(100) and C(5-7) layers, respectively. The black parallelepiped represents the unit cell spanned by the lattice vectors, ${\vec{a}_1}$, ${\vec{a}_2}$, and ${\vec{a}_3}$.}\label{fig1}
\end{figure*}

From the extensive conformational search, we obtained many direct gap carbon allotropes with $sp^{3}$-hybridized bonds.
When visually inspected, we notice that some of the low-energy ones are represented by the superlattice structure of C(100)$_n$/C(5-7), where $n$ C(100) layers ($n$ = 2, 3, and 4 for $N$ = 12, 16, and 20, respectively) and a defective layer denoted as C(5-7) are stacked in an alternating fashion (Fig.~\ref{fig1}).
The C(100) layers are almost identical to their corresponding layers of diamond along the [100] direction (here along the $z$-axis).
The C(5-7) and its neighboring C(100) layers form alternating 5- and 7-membered rings along the $y$-axis.
Recently, we reported Si superlattices with direct and quasidirect band gaps \cite{Oh2015}, where multiple Si(111) layers and a defective layer containing Seiwatz chains are stacked in a similar fashion as in C(100)$_n$/C(5-7).
In the Si superlattices, however, the Si layers are oriented along the [111] direction and the 5-8-type rings are formed in the defective region.

For superlattices with odd $n$, we found the same Bravais lattice, simple monoclinic, with the space group $P2/m$ (Table~\ref{t1}).
On the other hand, two types of superlattices (referred to as type-I and type-II) exist for even $n$, depending on the relative positions of defective layers.
As shown in Fig.~\ref{fig1}, the ${\vec{a}_3}$ vectors differ by $0.5\vec{a}_2$ between type-I and type-II.
The type-I (type-II) Bravais lattice is a base-centered (body-centered) orthorhombic lattice, with the space group of $Cmcm$ ($Imma$).
We note that 6 out of 10 superlattices in Table~\ref{t1} were previously reported separately \cite{P2012,H2012,Z4A3B1-2012,C2012,F2012,J2012}.
However, no systematic understanding of the structural variations of the superlattices in terms of $n$ and the dependence of Bravais lattice on the defective layers was attempted previously.
Here we explicitly investigate the optical performance of these superlattices in a systematic way. 

\begin{table*}[ht]
\caption{Summary of the structural and electronic properties of the C(100)$_n$/C(5-7) superlattices. For each structure, the number of atoms per unit cell ($N$), Bravais lattice, volume per atom ($\Omega_c$ in \AA$^3$/atom), total energy ($E_{tot}$ in meV/atom) relative to cubic diamond, direct (D) and indirect (ID) band gap sizes ($E_g$ in eV), and space group are shown. Here $E_g$(PBE) and $E_g$($G_0W_0$) denote the band gaps from the PBE exchange-correlation functional and quasiparticle $G_0W_0$ calculations, respectively. Bravais lattices are abbreviated, such as sm: simple monoclinic, base-co: base-centered orthorhombic, body-co: body-centered orthorhombic. Note that the S-S$_1$Z$_{2m}$ allotropes \cite{P2012} are identical to our C(100)$_{2m+1}$/C(5-7) superlattices.}\label{t1}
\begin{tabular}{lcccccccc}
\hline
\multicolumn{1}{c}{$n$(type)} & $N$ & lattice &  $\Omega_c$ & $E_{tot}$ & $E_g$(PBE) & $E_g$($G_0W_0$) &  space group & previous study \\  \hline
1          & 8  & sm      & 5.99 & 164 & 4.358(ID)  & 5.550(ID)  & $P2/m$ (No. 10) & F carbon \cite{F2012}, J carbon \cite{J2012} \\
2(I)  & 12 & base-co & 5.88 &  90 & 4.311(D)      & 5.936(D)      & $Cmcm$ (No. 63) & S carbon \cite{H2012}, C carbon \cite{C2012}\\
2(II) & 12 & body-co & 5.92 & 134 & 4.202(D)      & 5.804(D)      & $Imma$ (No. 74) & Z-CACB \cite{Z4A3B1-2012} \\
3          & 16 & sm      & 5.85 &  88 & 4.163(D)      & 5.764(D)      & $P2/m$ (No. 10) & S-S$_1$Z$_2$ \cite{P2012}\\
4(I)  & 20 & base-co & 5.83 &  73 & 4.091(D)      & 5.679(D)      & $Cmcm$ (No. 63) \\
4(II) & 20 & body-co & 5.82 &  70 & 4.139(D)      & 5.731(D)      & $Imma$ (No. 74) \\
5          & 24 & sm      & 5.81 &  61 & 4.076(D)      & 5.657(D)      & $P2/m$ (No. 10) & S-S$_1$Z$_4$ \cite{P2012} \\
6(I)  & 28 & base-co & 5.79 &  53 & 4.042(D)      & 5.614(D)      & $Cmcm$ (No. 63) \\
6(II) & 28 & body-co & 5.79 &  53 & 4.034(D)      & 5.606(D)      & $Imma$ (No. 74) \\
7          & 32 & sm      & 5.78 &  46 & 4.001(D)      & 5.569(D)      & $P2/m$ (No. 10) & S-S$_1$Z$_6$ \cite{P2012} \\    \hline
\end{tabular} 
\end{table*}

We examined the band structures of C(100)$_n$/C(5-7) for $n$ up to 10 using the PBE functional (Fig.~\ref{fig2} and see Appendix A).
The superlattice with $n$ = 1 shows an indirect band gap of about 4.36 eV.
For $n \ge 2$, all the superlattices are of direct band gaps located at the $\Gamma$ point, which were estimated to be 4.00$\sim$4.31 eV for $n$ = 2$\sim$7 (Table~\ref{t1}).
When more accurate $G_0W_0$ calculations \cite{GW1} were performed, the gap sizes were improved to 5.57$\sim$5.94 eV, without changing the nature of band gaps.
We note that the indirect and direct band gaps of diamond estimated by $G_0W_0$ are 5.45 and 7.34 eV, respectively, in good agreement with their experimental values of 5.5 and 7.1 eV \cite{expbandgap1,expbandgap2}.
We also note that both the equilibrium volumes and the direct band gaps tend to decrease with increasing $n$. 

\begin{figure}
\includegraphics[width=0.95\columnwidth]{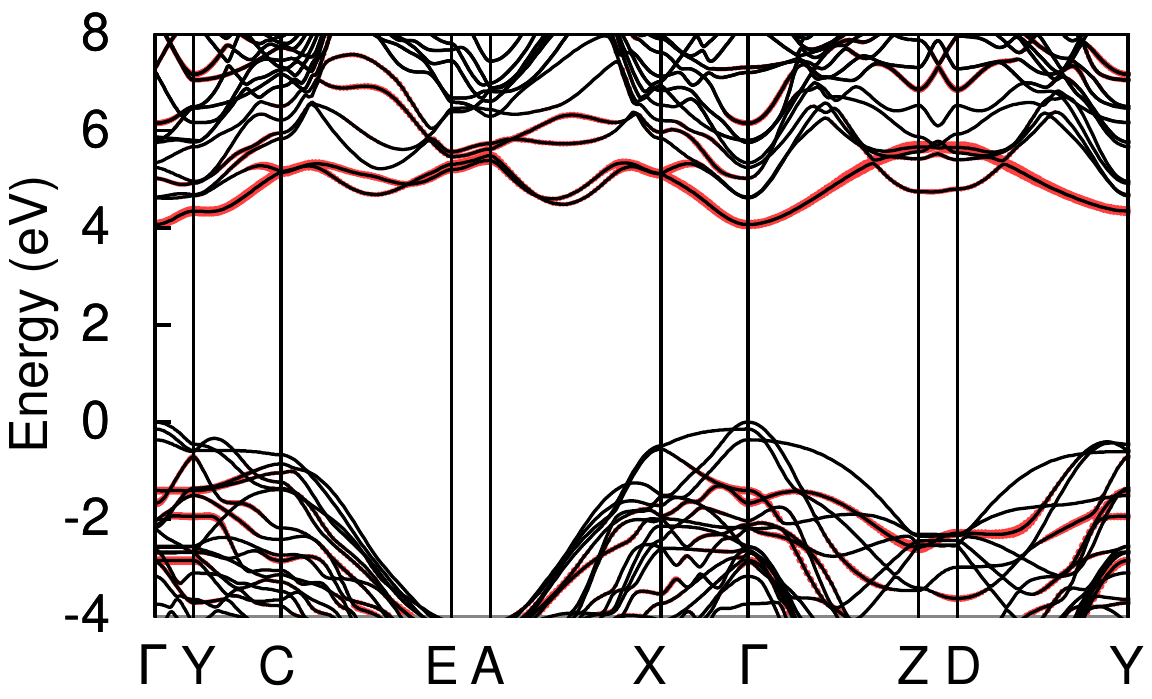}
\caption{The PBE band structure of the C(100)$_{n=5}$/C(5-7) superlattice with the valence band maximum set to zero. The thickness of red colored bands represents the degree of confinement in the defective region.}\label{fig2}
\end{figure}

\begin{figure}
\includegraphics[width=0.95\columnwidth]{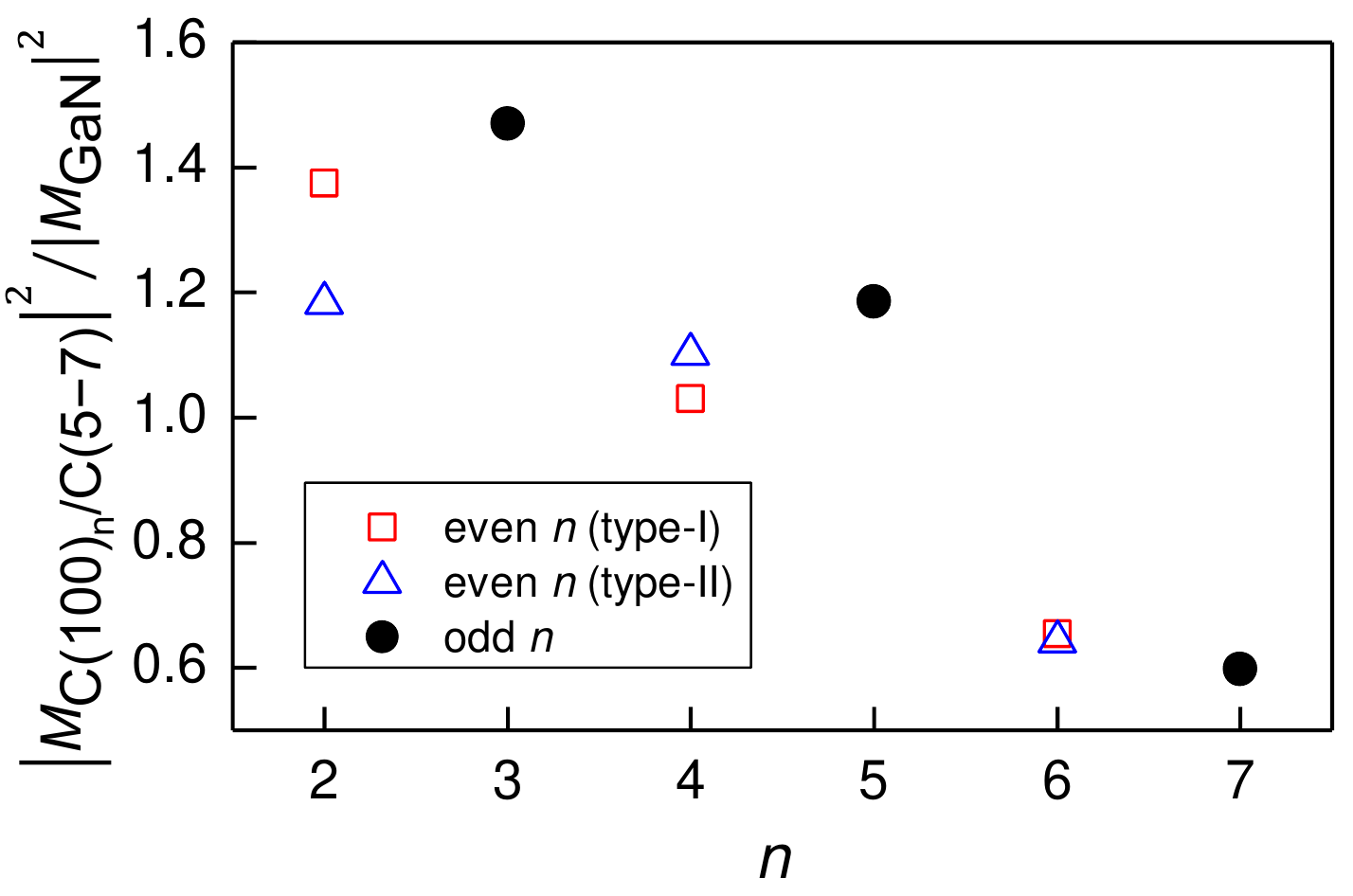}
\caption{The squares of the dipole matrix elements, $\vert M_{\textrm{C}(100){_n}/\textrm{C}(5-7)} \vert^2$, for the optical transition at the threshold energy in the C(100)$_n$/C(5-7) superlattices are compared with that of GaN, $\vert M_{\textrm{GaN}}\vert^2$.}\label{fig3}
\end{figure}

For proper evaluation of a carbon allotrope as an optoelectronic device, it is important to investigate its optical properties, especially the dipole allowedness of the optical transition at the band edge.
According to the Fermi's golden rule, the optical transition probability at the threshold energy is proportional to the square of the dipole matrix element, $\vert M \vert^2=\vert \langle f \vert \vec{p} \vert i \rangle \vert^2$.
For the superlattices, we calculated $\vert M \vert^2$ by solving the Bethe-Salpeter equation \cite{BSE} together with the $G_0W_0$ approximation \cite{GW1}.
For $n$ = 2$\sim$5, the optical transitions are stronger than that of GaN (Fig.~\ref{fig3}).
For $n \ge 6$, they decrease with increasing $n$, but, their overall values are comparable to that of GaN, suggesting that our superlattices may be used as a deep UV light source.
Recently, many carbon allotropes were predicted to be direct band gap semiconductors \cite{T2011,H2012,Z4A3B1-2012,C24-2006,TY2012,sp2-2013,Amm2-2014}, including S carbon, C carbon, Z-CACB, Z-ACA, fullerite C$_{24}$, T carbon, TY carbon, $sp^2$-diamond, and $Amm$2 carbon.
As listed in Table~\ref{t1}, the first three allotropes are equivalent to our superlattices with $n$ = 2.
Our $G_0W_0$ calculations show that T carbon is of an indirect band gap of 2.86 eV, while its direct band gap is 3.69 eV.
Moreover, $sp^2$-diamond has a small direct band gap of about 3 eV and its optical transition measured by $\vert M \vert^2$ is only about 10\% of GaN's.
The rest of the allotropes have either high energies (0.17$\sim$1.29 eV/atom) or small band gaps less than 3 eV, which are not suitable for deep UV optoelectronics. 

\begin{figure}
\centering
\includegraphics[width=0.95\columnwidth]{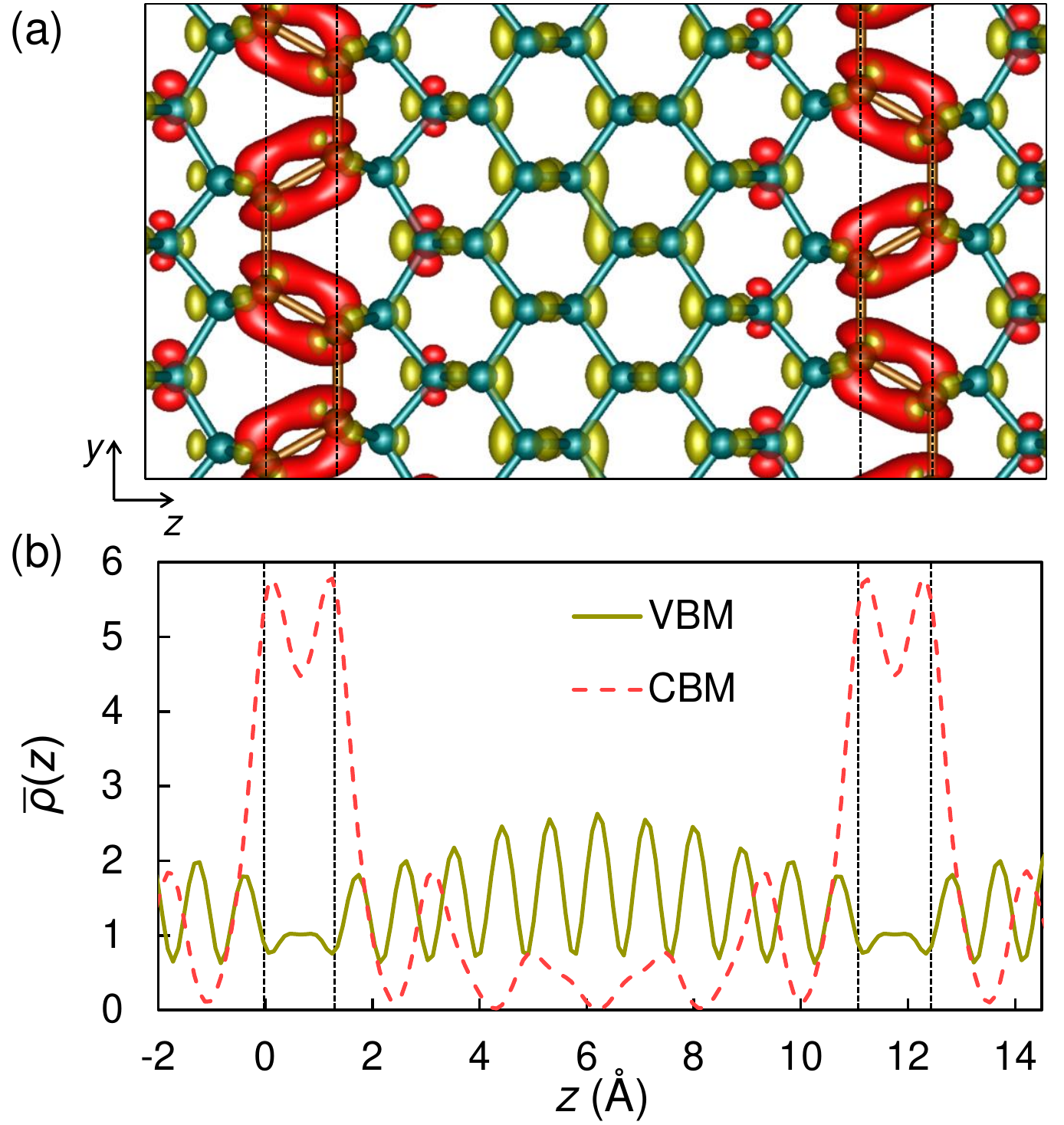}
\caption{For the C(100)$_{n=5}$/C(5-7) superlattice, (a) isosurfaces (0.07 electrons per {\AA}$^{3}$) of the charge densities of VBM (yellow) and CBM (red) and (b) their planar-averaged charge densities (in units of 10$^{-2}$ electrons per {\AA}$^{3}$) are plotted along the superlattice direction ($z$-axis). Black vertical lines denote the positions of the C atoms in the C(5-7) layers.}\label{fig4}
\end{figure}

To understand the strong optical transition, we analyzed the characteristics of the band edge states (Fig.~\ref{fig4}).
It is clear that the charge density of the valence band maximum (VBM) is preferentially distributed at the bonding sites of the C(100) layers as in diamond.
The planar-averaged charge densities also show that the bulk-like C(100) layers contribute to VBM more than the defective layers.
On the other hand, the charge density of the conduction band minimum (CBM) is largely confined in the defective layers, thus ruling out the possibility that the direct gap nature of our superlattices results from the effect of zone folding (see Appendix B).
In fact, the less dispersive band derived from the defective layers lies near the conduction band edge (Fig.~\ref{fig2}).
Due to the large overlap of the band edge states near the interface layers, dipole-allowed optical transitions are significantly enhanced in our superlattices.

\begin{figure}
\centering
\includegraphics[width=0.95\columnwidth]{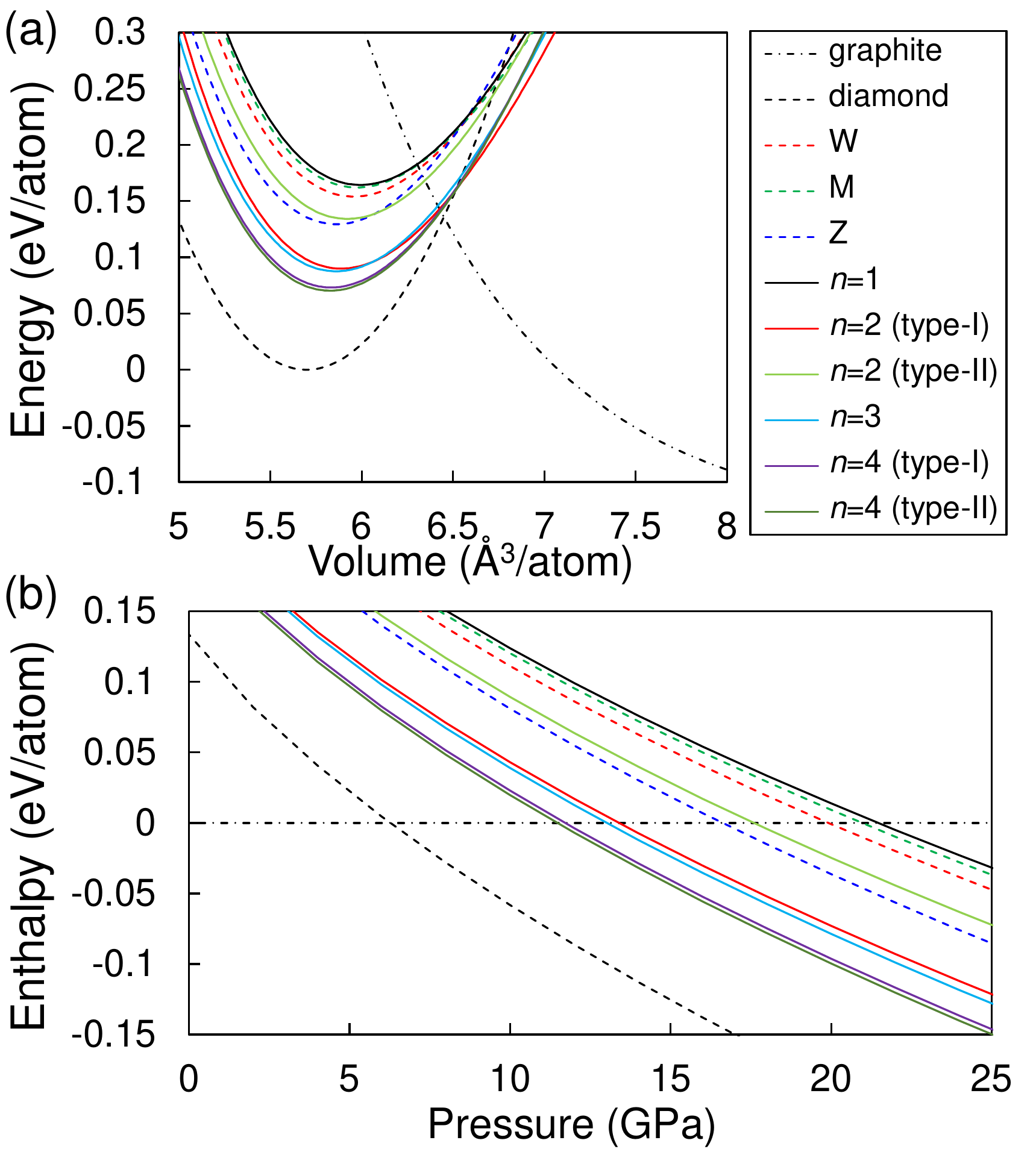}
\caption{The total energy versus volume and enthalpy versus pressure curves for various carbon allotropes \cite{M2009,W2011,Z2012} and the C(100)$_n$/C(5-7) superlattices are calculated by using the PBE functional.}\label{fig5}
\end{figure}

\begin{figure}
\centering
\includegraphics[width=0.95\columnwidth]{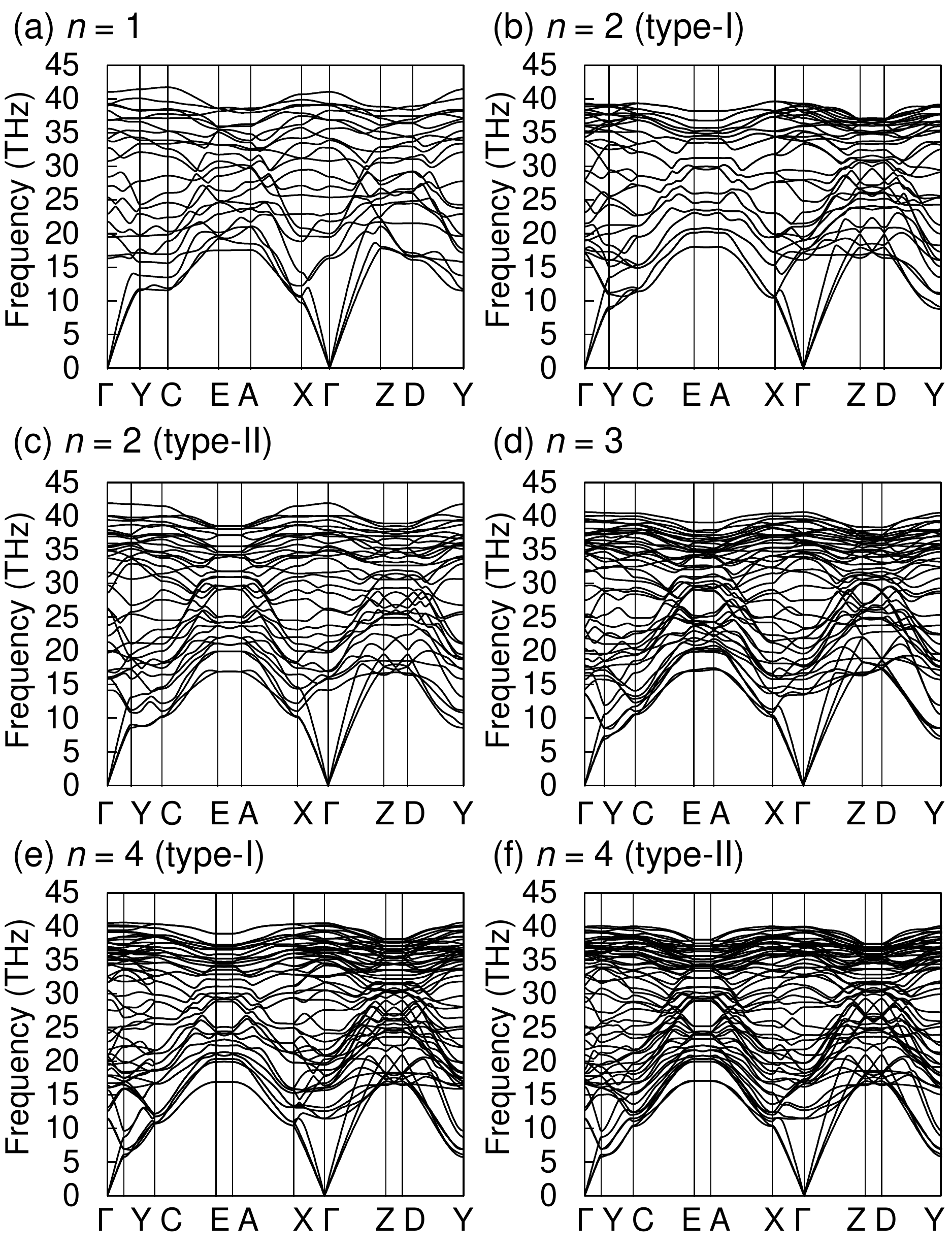}
\caption{The full phonon spectra of the C(100)$_n$/C(5-7) superlattices with $n = 1\sim4$. The dynamical matrices are calculated by using the $4\times3\times3$, $4\times3\times3$, $4\times3\times2$, and $4\times3\times2$ supercells for $n = 1$, 2, 3, and 4, respectively.}\label{fig6}
\end{figure}

\begin{figure}
\centering
\includegraphics[width=0.95\columnwidth]{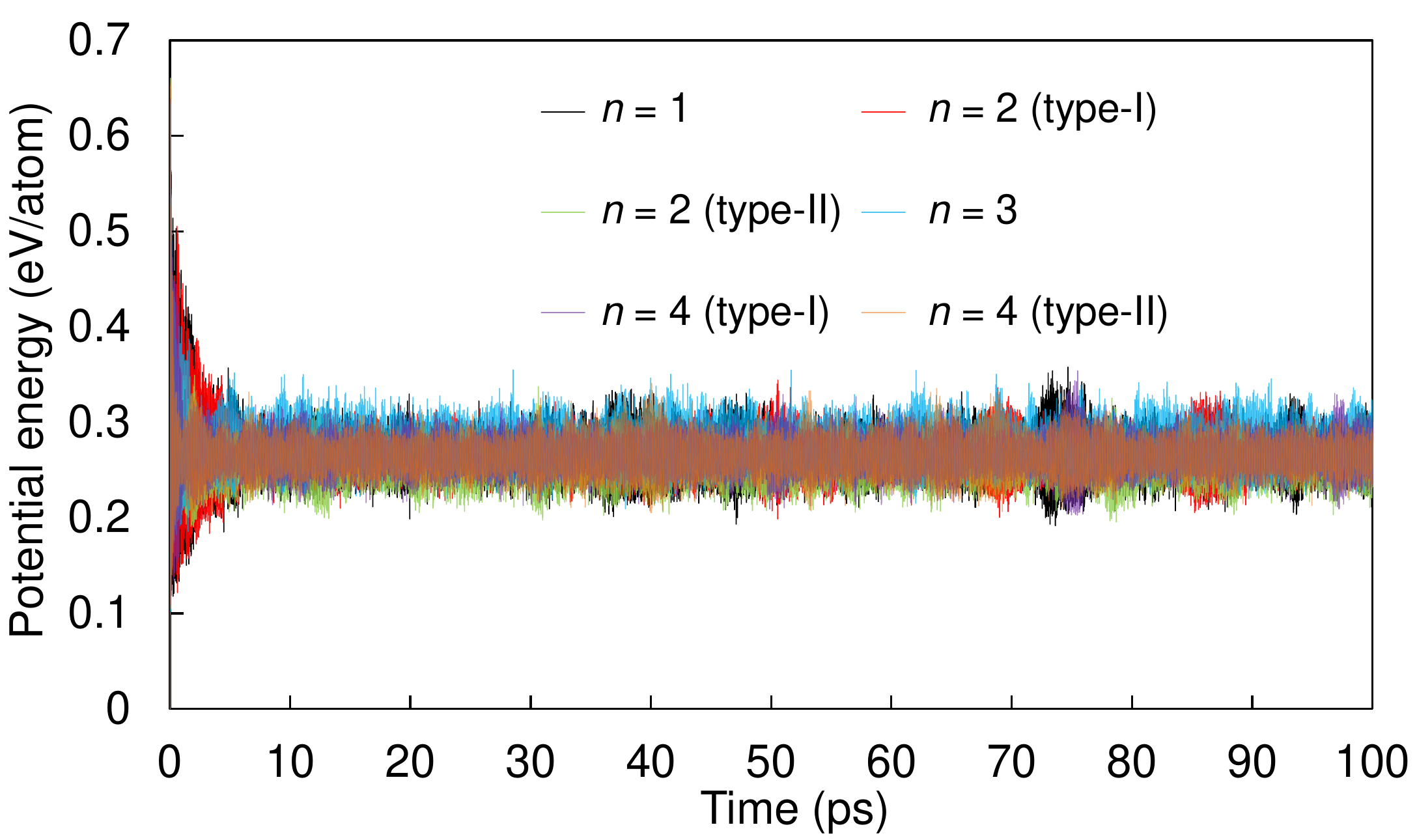}
\caption{The variations of potential energies during finite-temperature {\it {ab initio}} molecular dynamics simulations at 2000 K are shown for the C(100)$_n$/C(5-7) superlattices with $n = 1\sim4$. The $4\times2\times3$, $4\times2\times3$, $4\times2\times1$, and $4\times2\times1$ supercells are chosen for $n = 1$, 2, 3, and 4, respectively.}\label{fig7}
\end{figure}

Since the stacking of the C(100) layers is identical to that of diamond, the enthalpy of the superlattice tends to decrease with increasing $n$.
The excess energies of the superlattices with $n \ge 3$ are less than 90 meV per atom (Table~\ref{t1}), and these energies are lower than those of the previously reported non-superlattice allotropes (Fig.~\ref{fig5}).
When we examined the dynamical stability by calculating the full phonon spectra, we found no imaginary phonon modes (Fig.~\ref{fig6}).
In addition, we carried out first-principles molecular dynamics simulations and confirmed that all the superlattices were stable up to 100 ps at the high temperature of 2000 K (Fig.~\ref{fig7}).
Owing to the high thermal stability, the superlattices can be used for high power devices.

\begin{figure}
\centering
\includegraphics[width=0.95\columnwidth]{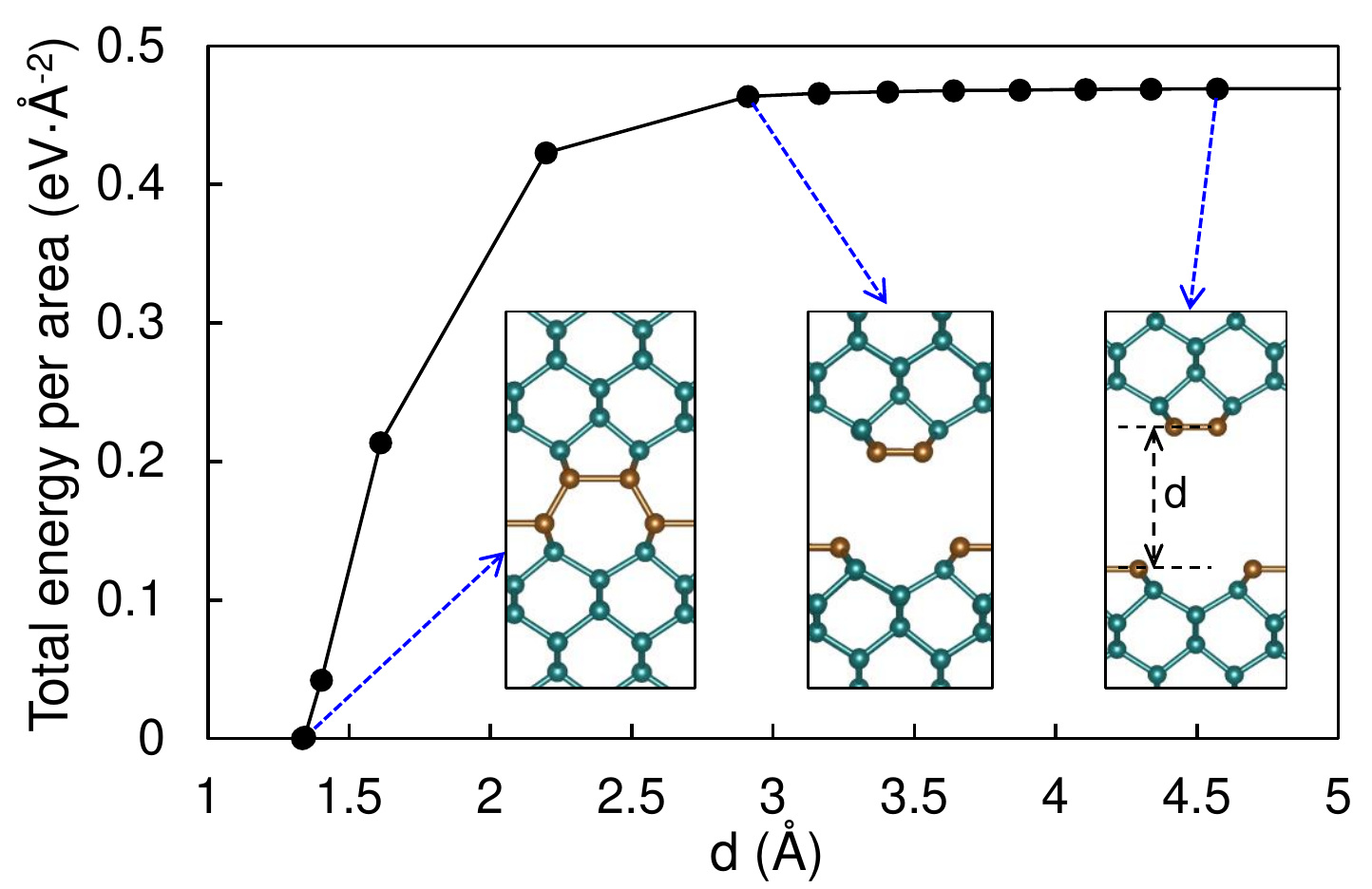}
\caption{The variation of total energy is plotted as a function of the distance between two clean C(100) surfaces.}\label{fig8}
\end{figure}

We propose that the defective layer can be formed from wafer bonding between two C(100) $2\times1$ surfaces, which can lead to the realization of our superlattice structures.
Atomically smooth C(100) surfaces can be prepared through the growth of homoepitaxial layers by chemical vapor deposition \cite{c100-1,c100-2}.
Upon thermal annealing, hydrogen desorption takes place and leads to the formation of clean C(100) $2\times1$ surfaces via surface reconstruction \cite{c100-3,c100-4,c100-5,c100-6}.
Figure~\ref{fig8} shows the variation of total energy during the process of wafer bonding between two clean C(100) $2\times1$ surfaces.
As the distance between the two surfaces decreases, the wafer bonding proceeds without any energy barrier, resulting in the C(5-7) defective layer.
In the case of C(100) $2\times1$ monohydride surfaces, H desorption must occur via the formation of H$_2$ molecules.
However, since this process is endothermic with a large energy barrier (at least 0.87 eV$\cdot${\AA}$^{-2}$), the C(5-7) layer cannot be easily formed.

Finally, we calculated the effective masses of hole and electron carriers along the principal axes from the $\Gamma$ point, denoted as $m_h$ and $m_e$, respectively (Table~\ref{t2}).
For $n$ = 2$\sim$10, the average values of $m_h$ lie in the range of 0.34$\sim$0.39 $m_{0}$, nearly independent of $n$, where $m_{0}$ is the bare electron mass, and these effective masses are comparable to that (0.43 $m_{0}$) of diamond.
On the other hand, the electron effective masses are more anisotropic than for holes, similar to the case of diamond.
Since the lowest conduction band originates from the C(5-7) layers, its band dispersion becomes weakened as the distance between adjacent defective layers increases.
As a consequence, the electron effective mass increases as $n$ increases.
The average values of $m_e$ are estimated to be 0.53$\sim$1.42 $m_{0}$.
For small $n$, the electron effective masses are similar to that (0.40 $m_{0}$) of diamond, however, they increase by a factor of two or three for large $n$.
Thus, short-period superlattices with small $n$ are desirable to obtain high carrier mobility for both electrons and holes.
 
\begin{table}
\caption{The effective masses of electron and hole carriers in the C(100)$_n$/C(5-7) superlattices.
Three principal values of the effective mass tensor of holes (electrons) are denoted as $m_{h_1}$, $m_{h_2}$, and $m_{h_3}$ ($m_{e_1}$, $m_{e_2}$, and $m_{e_3}$) in units of the bare electron mass ($m_0$).
The harmonic mean value of $m_{h_1}$, $m_{h_2}$, and $m_{h_3}$ is given by $\bar{m}_{h}$, whereas that of $m_{e_1}$, $m_{e_2}$, and $m_{e_3}$ by $\bar{m}_{e}$.
The PBE functional for the exchange-correlation potential is used, and the spin-orbit coupling is included.
The calculated effective mass of holes in cubic diamond is found to be 0.43 $m_0$.
The longitudinal and transverse masses of electrons are calculated to be 0.29 $m_0$ and 1.61 $m_0$, respectively, in good agreement with the experimental values of 0.28 $m_0$ and 1.56 $m_0$ \cite{emass}.}\label{t2}
\begin{tabular}{lcccccccc}
\hline
\multicolumn{1}{c}{$n$(type)} & $m_{h_1}$ & $m_{h_2}$ & $m_{h_3}$ & $\bar{m}_{h}$ & $m_{e_1}$ & $m_{e_2}$ & $m_{e_3}$ & $\bar{m}_{e}$  \\ \hline
2(I) &0.27 &0.43 &0.55 &0.38 &0.38 &0.50 &1.08 &0.54 \\
2(II) &0.30 &0.42 &0.50 &0.39 &0.39 &0.47 &1.12 &0.54 \\
3 &0.29 &0.40 &0.54 &0.38 &0.39 &0.45 &1.14 &0.53 \\
4(I) &0.29 &0.36 &0.51 &0.37 &0.43 &0.48 &1.05 &0.56 \\
4(II) &0.28 &0.36 &0.51 &0.36 &0.43 &0.48 &1.00 &0.55 \\
5 &0.28 &0.37 &0.57 &0.37 &0.53 &0.67 &1.34 &0.73 \\
6(I) &0.28 &0.35 &0.56 &0.37 &0.59 &0.78 &1.24 &0.79 \\
6(II) &0.28 &0.35 &0.55 &0.36 &0.59 &0.78 &1.26 &0.80 \\
7 &0.27 &0.38 &0.58 &0.37 &0.73 &0.84 &1.21 &0.89 \\ 
8(I) &0.25 &0.35 &0.51 &0.34 &1.02 &1.03 &1.34 &1.11 \\ 
8(II) &0.25 &0.35 &0.52 &0.34 &1.01 &1.01 &1.29 &1.09 \\
9 &0.26 &0.35 &0.51 &0.35 &1.07 &1.33 &1.39 &1.25 \\
10(I) &0.27 &0.33 &0.52 &0.35 &1.12 &1.35 &1.89 &1.39 \\
10(II) &0.27 &0.33 &0.52 &0.35 &1.16 &1.35 &1.97 &1.42 \\ \hline
\end{tabular} 
\end{table}

\section{Conclusions}
In conclusion, we have predicted a family of pure C-based superlattices that are of wide and direct band gaps.
These superlattices can serve as a light emitter in the range of 210$\sim$221 nm, with efficiencies comparable to that of GaN.
The superlattice structure is characterized by alternating stacks of diamond (100) layers and a defective layer which forms the 5- and 7-membered rings with its neighboring layers.
Since a large portion of superlattices comes from the diamond structure, their energies are quite low and consequently their stabilities are greatly enhanced.
With the excellent thermal stability, the carbon superlattices have the potential to be used in optoelectronic devices which operate at very high voltage and temperature. 
We have shown that the defective layer can be formed through wafer bonding of two C(100) surfaces,
which can lead to the realization of the superlattice structure.
Our finding of direct gap carbon superlattices opens up the possibility of utilizing carbon allotropes for optoelectronic applications in a deep ultraviolet wavelength range.

\section*{ACKNOWLEDGEMENTS}
YJO, SK, IHL, and KJC were supported by Samsung Science and Technology Foundation under Grant No. SSTF-BA1401-08.
IHL and JL were supported by the National Research Foundation of Korea (NRF) under Grant No. 2008-0061987 funded by the Korea government.
We thank Korea Institute for Advanced Study (KIAS Center for Advanced Computation)
for providing computing resources.

\appendix

\section{LATTICE VECTORS AND HIGH SYMMETRY POINTS IN BRILLOUIN ZONE}
In the C(100)$_n$/C(5-7) superlattices, the Bravais lattice is determined by the number of the C(100) layers and the relative positions of the C(5-7) defective layers, as shown in Table~\ref{t1}.
To compare consistently the band structures of these superlattices in the same type of Brillouin zone (BZ), we choose the lattice vectors, $\vec{a}_1$, $\vec{a}_2$, and $\vec{a}_3$, which are defined as
$\vec{a}_1=\left( a/\sqrt{2}, 0, 0 \right) $,
$\vec{a}_2=\left( 0, \sqrt{2}a, 0 \right)$,
$\vec{a}_3=\left( ia/2\sqrt{2}, ja/2\sqrt{2}, 2.16 + na/2 \right) $,
where $a$ is the lattice constant of cubic diamond in {\AA} unit.
For the superlattices with odd $n$, $i=0$ and $j=1$.
In the case of even $n$, $i=1$ and $j=0$ ($2$) for type-I (type-II) superlattices.
Since different structural relaxations occur in the superlattices, the actual lengths of the lattice vectors vary to within 3.0$\%$ for $n  \geq 2$.

\begin{figure}
\centering
\includegraphics[width=0.95\columnwidth]{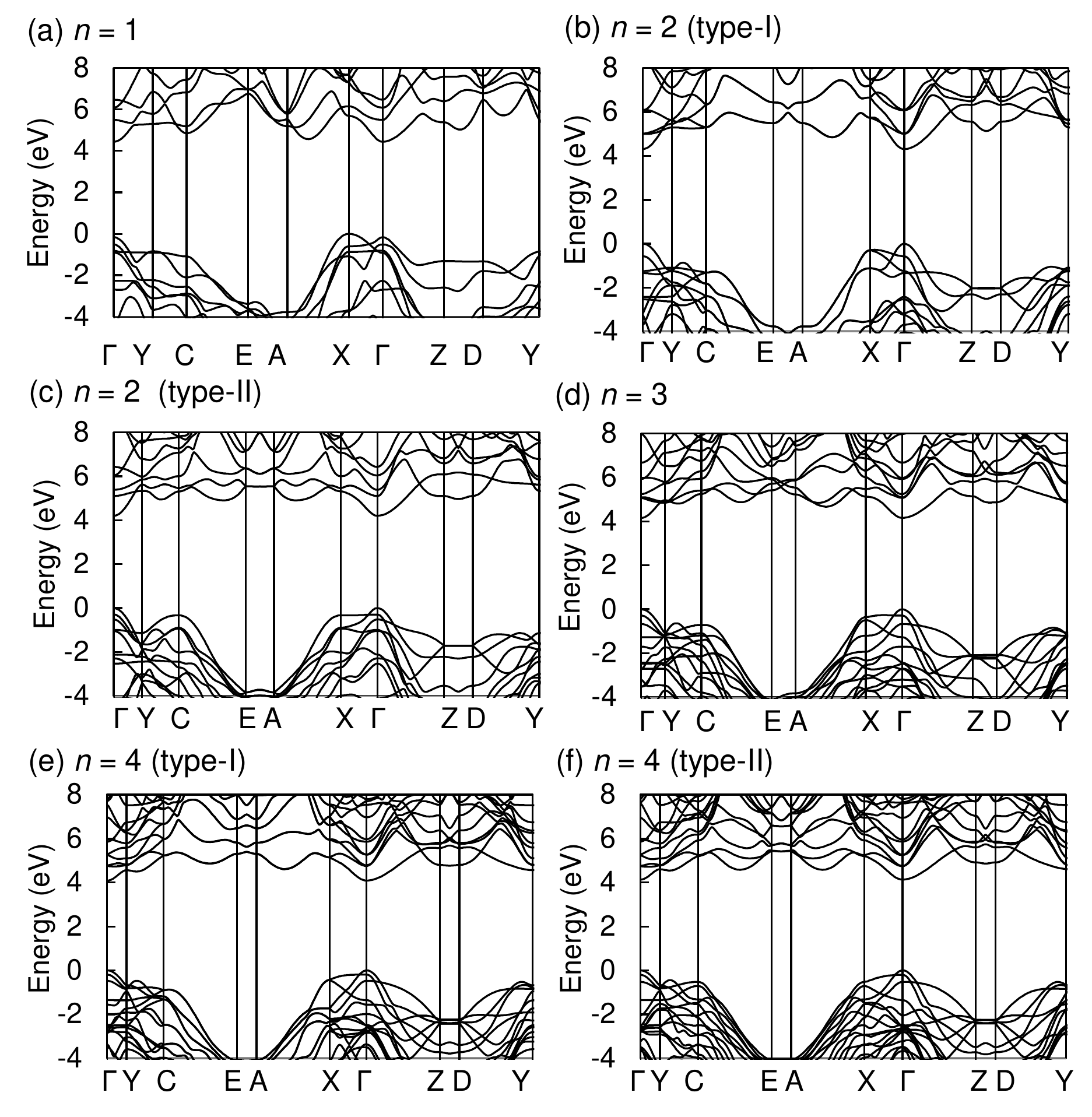}
\caption{The calculated band structures of the C(100)$_n$/C(5-7) superlattices with $n = 1\sim4$ by using the PBE functional. The energy of the valence band maximum is set to zero.}\label{fig9}
\end{figure}

\begin{figure}
\centering
\includegraphics[width=0.95\columnwidth]{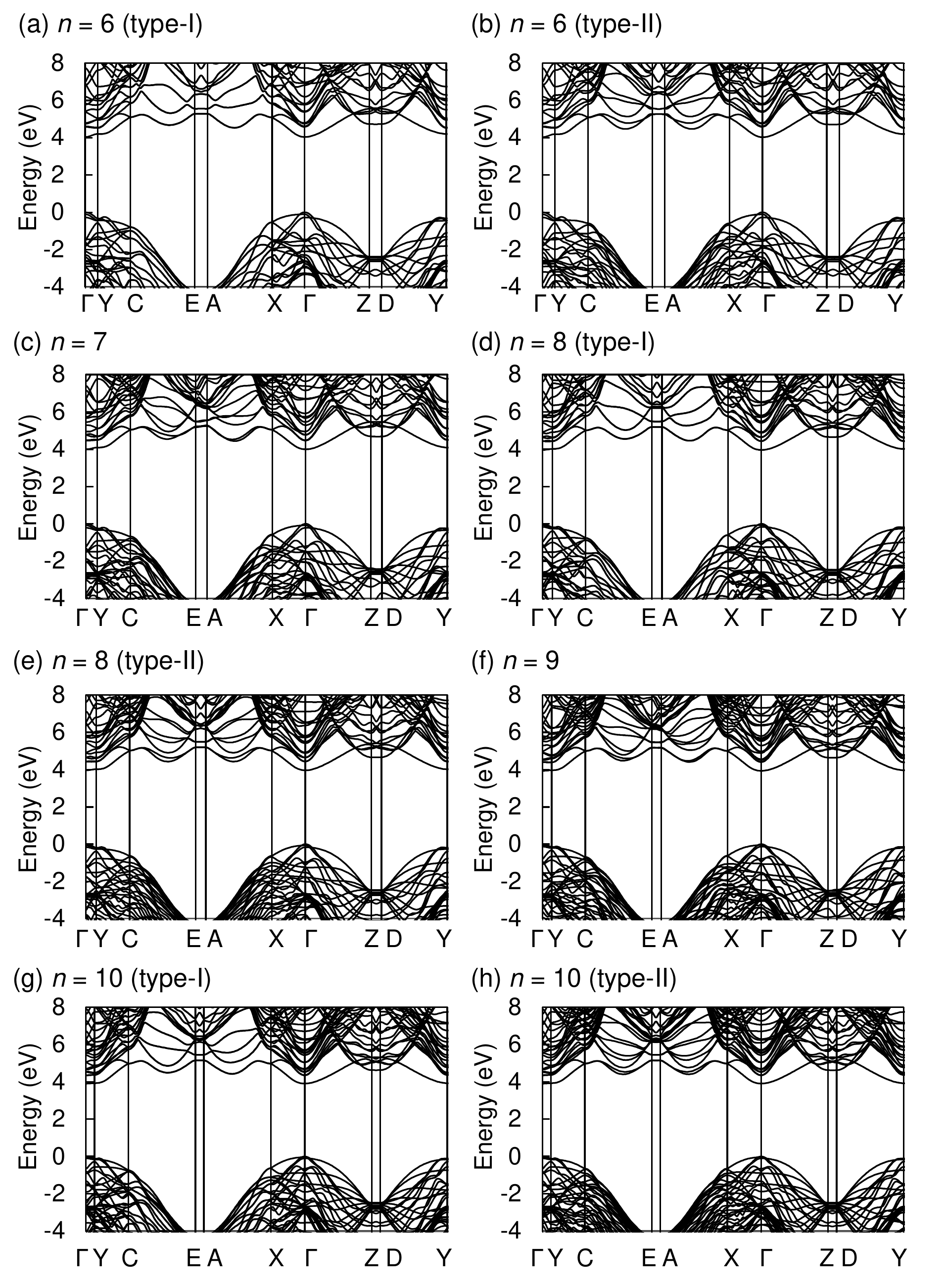}
\caption{The calculated band structures of the C(100)$_n$/C(5-7) superlattices with $n = 6\sim10$ by using the PBE functional. The energy of the valence band maximum is set to zero.}\label{fig10}
\end{figure}

The reciprocal lattice vectors, $\vec{b}_1$, $\vec{b}_2$, and $\vec{b}_3$, are determined from the lattice vectors as described above.
To draw the band structures of the C(100)$_n$/C(5-7) superlattices, the following notations are used for the high symmetry $\vec{k}$-points in the BZ:
$\Gamma = \left( 0, 0, 0 \right)$,
$\textrm{Y} = \left( 0, 0, 0.5 \right)$,
$\textrm{C} = \left( 0, 0.5, 0.5 \right)$,
$\textrm{E} = \left( 0.5, 0.5, 0.5 \right)$,
$\textrm{A} = \left( 0.5, 0.5, 0 \right)$,
$\textrm{X} = \left( 0, 0.5, 0 \right)$,
$\textrm{Z} = \left( 0.5, 0, 0 \right)$,
$\textrm{D} = \left( 0.5, 0, 0.5 \right)$,
where the $\vec{k}$-points are given in fractions of the reciprocal lattice vectors.
Note that, for even $n$, the lattice vectors, the reciprocal lattice vectors, and the notations for the high symmetry $\vec{k}$-points do not follow the conventional ones \cite{Satyawan}.
On the other hand, for odd $n$, the above notations are consistent with the conventional ones for the simple monoclinic lattice.
The visual comparison of the band structures is shown in Figs.~\ref{fig9} and~\ref{fig10}.

\section{ORBITAL CHARACTERISTICS OF BAND EDGE STATES}
To describe the orbital characteristics of the VBM and CBM states in the C(100)$_n$/C(5-7) superlattices with $n \geq 2$, we use the band unfolding scheme \cite{unfold1,unfold2,unfold3}.
As illustrated in Fig.~\ref{fig4}, the wave functions for the band edge states are distributed in both the bulk-like C(100) and defective C(5-7) layers.
Here we only consider the wave functions of VBM and CBM which are distributed in the C(100) layers, not in the defective layers.

First, we set up an artificial supercell which is made by repeating the $n$ C(100) layers but removing the defective layer in the unit cell of the C(100)$_n$/C(5-7) superlattice.
Second, we obtain the projected wave function by decomposing the confined wave function in the C(100) layers into the $l = 0$, $1$, and $2$ spherical harmonics.
In this case, we choose a sphere with the radius of 0.8 {\AA} around each C atom in the C(100) layers. 
Third, we determine a sampling of $\vec{k}^{\textrm{fcc}}$-points by using the folding relation, $\vec{k}^{\textrm{SC}}-\vec{G}^{\textrm{SC}}=\vec{k}^{\textrm{fcc}}$, based on the unfolding scheme, where $\vec{k}^{\textrm{fcc}}$ represents the $\vec{k}$-points in the first BZ of the fcc lattice of cubic diamond, whereas $\vec{k}^{\textrm{SC}}$ for those in the first BZ of the artificial supercell.

\begin{figure}[ht]
\centering
\includegraphics[width=0.95\columnwidth]{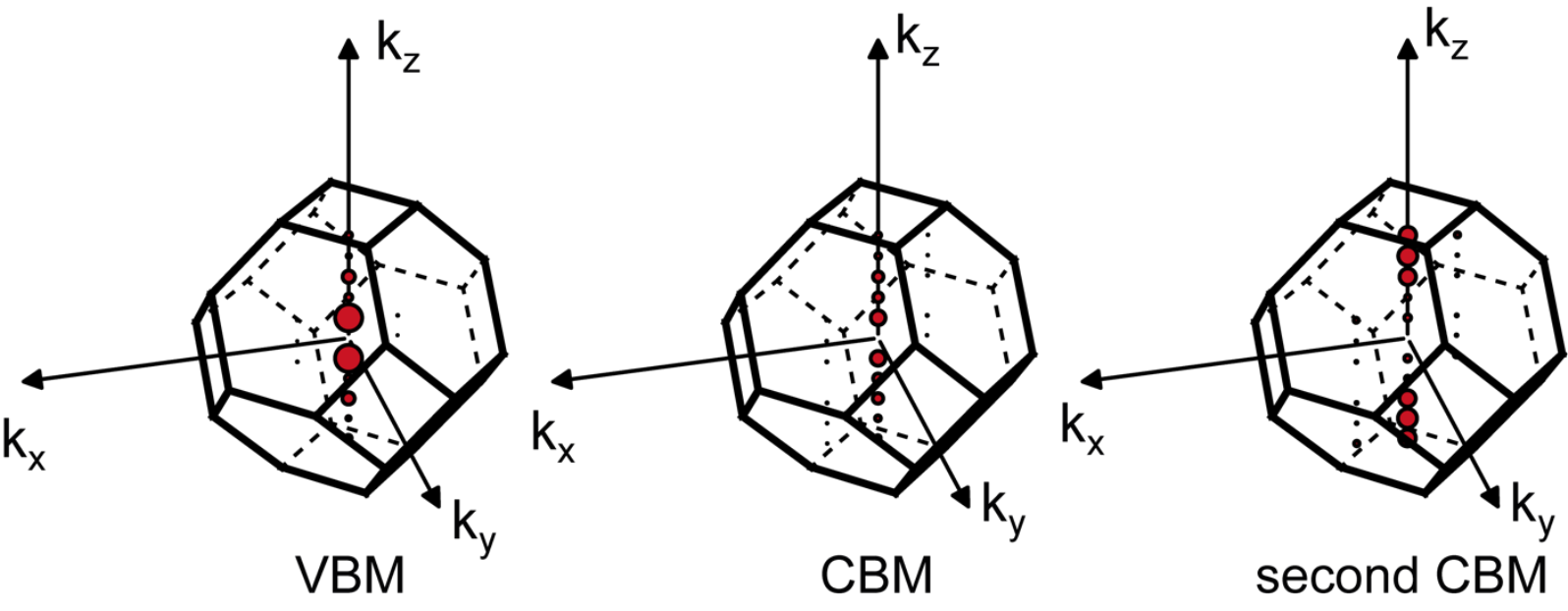}
\caption{The orbital characteristics of the VBM, CBM, and second CBM states at the $\Gamma$ point of the C(100)$_{n=5}$/C(5-7) superlattice. Black solid lines indicate the first BZ of the fcc lattice of cubic diamond. The size of red circles is proportional to the degree of contributions from the bulk states in the fcc BZ of cubic diamond. Note that the wave function of VBM is mostly distributed in the C(100) layers, and its character is mainly derived from the bulk states around the $\Gamma$ point. While the wave function of CBM is mostly confined in the defective layers, a small portion of its wave function is also distributed in the C(100) layers, and its orbital character is similar to those for the bulk states around the $\Gamma$ point. On the other hand, the second CBM state is contributed from the bulk states around the X-valley and thus attributed to the zone folding effect.}\label{fig11}
\end{figure}

Here, $\vec{k}^{\textrm{SC}}$ = (0, 0, 0), because the orbital characteristics of VBM and CBM in the C(100)$_n$/C(5-7) superlattices are analyzed at the $\Gamma$ point.
The $k_x$ and $k_y$ components of $\vec{G}^{\textrm{SC}}$ are given in terms of the reciprocal vectors on the lateral plane, $\frac{2\pi}{a}(\sqrt{2}, 0, 0)$ and $\frac{2\pi}{a}(0, 1/\sqrt{2}, 0)$, where $a$ is the lattice constant of cubic diamond. 
Along the $k_z$ direction, instead of taking into account the periodicity of the supercell, we consider an isolated slab which only consists of the $n$ C(100) layers. 
Then, the wave function along the direction perpendicular to the lateral plane is described by a confined standing wave instead of the folded Bloch wave \cite{unfold1}. 
Using the H\"uckel theory \cite{unfold4} for the confined standing wave, we obtain the $k_z$ component of $\vec{G}^{\textrm{SC}}$ which satisfies the relation, $\vec{G}_{k_z}^{\textrm{SC}}=\pm \frac{2\pi}{a}  \frac{m}{n+1} \hat{k}_z$, where $m$ is a positive integer.
Note that, for $m > n$, $\vec{G}^{\textrm{SC}}$ is located outside the first BZ of cubic diamond. 
Thus, we choose the $m$ values ($m = 1$, $2$, $\ldots$, $n$) for the $\vec{k}^{\textrm{fcc}}$ sampling. 
For example, in the C(100)$_n$/C(5-7) superlattice with $n = 5$, five $\vec{k}^{\textrm{fcc}}$ points are chosen along the $\Gamma$--X line, as shown in Fig.~\ref{fig11}.

\section{STRUCTURAL INFORMATION ON THE SUPERLATTICES}
The structural information on the lattice parameters and Wyckoff positions of the C atoms are shown in Tables~\ref{t3} and~\ref{t4}.

\begin{table}
\caption{Lattice parameters of the C(100)$_n$/C(5-7) superlattices with $1 \leq n \leq 5$.}\label{t3}
\begin{tabular}{lcccc}
\hline
\multicolumn{1}{c}{$n$(type)} & $a$({\AA}) & $b$({\AA}) & $c$({\AA}) & $\beta$($^\circ$) \\ \hline
1  & 4.13 & 2.53 & 4.78 & 106.07 \\
2(I)  & 2.52 & 11.41 & 4.91 & 90.00 \\
2(II)  & 2.53 & 4.87 & 11.53 & 90.00 \\
3  & 4.93 & 2.52 & 7.62 & 99.32 \\
4(I)  & 2.52 & 18.62 & 4.96 & 90.00 \\
4(II)  & 2.52 & 4.96 & 18.60 & 90.00 \\
5  & 4.98 & 2.52 & 11.16 & 96.40 \\ \hline
\end{tabular}
\end{table}

\newpage

\begin{table}
\caption{Wyckoff positions of the C atoms in the C(100)$_n$/C(5-7) superlattices with $1 \leq n \leq 5$.}\label{t4}
\begin{tiny}
\begin{tabular}{lcccc}
\hline
\multicolumn{1}{c}{$n$(type)} & site & \multicolumn{3}{c}{fractional coordinates} \\ \hline
1& 2m &0.17493& 0.00000& 0.11755\\
& 2m &0.16679& 0.00000& 0.46604\\
& 2n &0.63152& 0.50000& 0.42102\\
& 2n &0.60016& 0.50000& 0.88665\\ \hline
2(I)&8f &0.00000& -0.05831& -0.07833\\ %
&8f &0.00000& 0.63217& 0.48208\\
&4c &0.00000& 0.29961& 0.25000\\
&4c &0.00000& 0.72163& 0.25000\\  \hline
2(II)&8h &0.00000& 0.57262& 0.05857\\ %
&8h &0.00000& 0.51262& 0.36801\\
&4e &0.00000& 0.25000& 0.29160\\
&4e &0.00000& 0.25000& 0.71411\\ \hline
3&2m &0.09772& 0.00000& 0.08874\\ %
&2m &0.44666& 0.00000& 0.08920\\
&2m &0.13915& 0.00000& 0.56593\\
&2m &0.64148& 0.00000& 0.55384\\ 
&2n &0.46574& 0.50000& 0.79791\\
&2n &-0.06568& 0.50000& 0.79948\\
&2n &0.17028& 0.50000& 0.68892\\
&2n &0.67014& 0.50000& 0.66810\\ \hline
4(I)&8f &0.00000& -0.03592& -0.07524\\ %
&8f &0.00000& 0.58130& 0.48396\\
&8f &0.00000& 0.22603& -0.00019\\ 
&4c &0.00000& 0.37419& 0.25000\\
&4c &0.00000& 0.63383& 0.25000\\
&4c &0.00000& 0.82435& 0.25000\\
&4c &0.00000& 0.17984& 0.25000\\  \hline
4(II)&8h &0.00000& 0.57587& 0.03590\\
&8h &0.00000& 0.51646& 0.41870\\
&8h &0.00000& 0.50536& 0.77400\\
&4e &0.00000& 0.25000& 0.37469\\
&4e &0.00000& 0.25000& 0.63428\\
&4e &0.00000& 0.25000& 0.17488\\
&4e &0.00000& 0.25000& 0.81939\\  \hline
5&2m &0.09063& 0.00000& 0.06023\\ %
&2m &0.43948& 0.00000& 0.06020\\
&2m &0.17668& 0.00000& 0.70605\\
&2m &0.67414& 0.00000& 0.69763\\
&2m &-0.09190& 0.00000& 0.62132\\
&2m &0.40234& 0.00000& 0.62050\\
&2n &0.48238& 0.50000& 0.86368\\
&2n &-0.05064& 0.50000& 0.86358\\
&2n &0.19756& 0.50000& 0.78946\\
&2n &0.69344& 0.50000& 0.77511\\
&2n &0.88790& 0.50000& 0.54106\\
&2n &0.38231& 0.50000& 0.53976\\ \hline
\end{tabular}
\end{tiny}
\end{table}

\end{document}